\documentclass[a4paper]{jpconf}
\usepackage{graphicx}
\begin{document}
\title{Self-Organizing Maps Algorithm for Parton Distribution Functions Extraction}

\author{Simonetta Liuti$^1$, Katherine A Holcomb$^2$, and Evan Askanazi$^1$}

\address{$^1$University of Virginia - Physics Department,  382, McCormick Rd., Charlottesville, VA 22904 - USA}
\address{$^2$University of Virginia - UVACSE,
112, Albert H. Small Building, Charlottesville, Virginia 22904 - USA} 

\ead{sl4y@virginia.edu, kah3f@virginia.edu, ema9u@virginia.edu}

\begin{abstract}
We describe a new method to extract parton distribution functions from hard scattering processes based on Self-Organizing Maps. The extension to a larger, and more complex  class of soft matrix elements, including generalized parton distributions is also discussed. 
\end{abstract}

\section{Introduction}
In the past twenty years Artificial Neural Networks (ANNs) have remarkably established their role as a computational tool in high energy physics analyses. 
Important applications have been developed that provide, for instance, classification methods for off-line jet identification and tracking, non-classification-type tools for on-line process control/event trigger and mass reconstruction, and optimization techniques in {\it e.g.}  track finding \cite{Pet}. 
More recently, ANNs have been proposed as a theoretical tool to address the problem of  the extraction of Parton Distribution Functions (PDFs) from high energy scattering processes. 
In a series of papers \cite{Ball.2010,DelDebbio.2007,Forte.2002} the authors developed the 
Neural Networks PDFs (NNPDFs).
NNPDFs differ  from standard global analyses in that they avoid the bias that is introduced when a  particular parametric functional form for the PDFs is adopted.  Relying on an initial PDF form is instead a fundamental assumption in standard analyses.  
In NNPDFs each PDF is  parameterized with a redundant functional form given by a neural network with 37 free parameters represented by the ANNs weights. 
The parametric  forms are subsequently
fitted to the experimental data, and their $\chi^2$ is minimized using a genetic algorithm \cite{DelDebbio.2007}. 

What distinguishes this approach from the standard ones is, besides an improved statistical analysis, the automated minimization procedure
which has the {\it advantage} of  eliminating the user's bias in extracting PDFs from experimental data. 
In fact, no physical assumption goes into determining the shape of the parametrization, rather the physical behavior is inferred directly from the data
in terms of smooth PDF curves. 
This feature turns, however,  into a {\it disadvantage} in  
kinematical regions where there is no experimental information, or in between the data points if the data are 
sparse: due to the inherent unbiasedness  NNPDFs 
cannot be sensibly extrapolated to kinematical regions with few or no data, starting from their behavior in regions where data do exist. 
In other words,  since for NNPDFs the effect of modifying individual
NN parameters is unknown -- the weights are in a non tunable, hidden layer --  the result might not be under control
in the extrapolation region.

The ability to extrapolate is, however, often a desirable feature in the performance of a fit to high energy physics data. It is even more 
important for extracting and processing physics information from a
new generation of experiments  that go beyond the inclusive scattering ones intensively studied so far.  
The newly accessible experiments include besides the LHC high energy frontier, semi-inclusive and exclusive scattering reactions with both unpolarized and polarized variables in the kinematical regimes accessible at RHIC and at Jefferson Lab$@$12 GeV. %Due to their exclusive nature, these measurements are sensibly more challenging than the inclusive-type ones, since they involve a larger number of observables that in turn depend on several kinematical variables. Thus their kinematical coverage has been determined so far spottily, and the observables have been measured with less accuracy.  

In the new LHC era it is of the utmost importance to provide methods to carefully extract the soft matrix elements, including PDFs and their uncertainties  with methods that are apt to handle and interpret the increasingly complicated, and diverse sets of observations. 
In this contribution we present an alternative approach based on Self-Organizing Maps (SOMs) \cite{Kohonen}. SOMs are also ANNs where, however,  the basic learning algorithm is replaced by  an {\em unsupervised learning} one that makes use of
the underlying {\em self-organizing} properties of the network's input vectors -- in our case the PDFs. Unsupervised learning
works without knowledge about what features characterize the output:
data are organized according to certain attributes without any ``teaching''. 
A most important aspect of the SOM algorithm is in its ability of projecting high dimensional {\it input} data onto lower dimensions representations while preserving the topological features present in the {\it training} data. 
Results using this technique can be represented as 2D geometrical configurations where one can visualize a neighborhood of nodes
being simultaneously updated while reproducing the clustering of the data's features. The 2D visualization features are characteristic of a map.   

In this contribution we present results obtained by restructuring the original SOMPDF code \cite{Carnahan} in such a way that we obtain directly smooth, 
continuous types of solutions for which a fully quantitative error analysis can be implemented.  Also, our new code is now sufficiently flexible to allow for analyses of different observables including both the unpolarized and polarized inclusive type structure functions, and the matrix elements for deeply virtual exclusive processes (Generalized Parton Distributions, GPDs), and semi-inclusive processes (Transverse Momentum Distributions, TMDs).
Our first quantitative results for the unpolarized case using Next-to-Leading-Order (NLO) perturbative QCD were shown in Refs.\cite{Perry_dis10,Hol_exc}. 
Here we present some of the intermediate results defining the working of SOMPDF.

\section{SOM Algorithm}
\label{sec2}
The SOM algorithm consists of three stages: {\it i) Initialization}; {\it ii) Training}; {\it iii) Mapping}. 
Each cell (neuron) is sensitized to a different domain of vectors.
 
\vspace{0.5cm}
\noindent 
{\it i) Initialization}

\noindent
During the initialization procedure weight vectors of dimension $n$ are associated to each cell, $i$,
\[ V_i = [v_i^{(1)}, ..., v_i^{(n)} ]  \] 
$V_i$ are given spatial coordinates, {\it i.e.} one defines the geometry/topology of a 2D map
that gets populated randomly with $V_i$.

\vspace{0.5cm}
\noindent 
{\it i) Training}

\noindent
For the training, a set of input data,
\[ \xi = [\xi^{(1)}, ..., \xi^{(n)} ] , \] 
(isomorphic to $V_i$) is then presented  to $V_i$, or compared by calculating the distance, $D_i$, between each pair of vectors via a ``similarity metric" that we choose to be $L_2$
(Euclidean distance) 
\[ D_i = \sqrt { \sum_{j=1,n} (v_i^{(j)} - \xi^{(j)})^2  } \]
The most similar weight vector is the Best Matching Unit (BMU). SOMs are based on  unsupervised and ``competitive" learning.
This means that the cells that are closest to the BMU activate each other in order to ``learn" from $\xi$.  Practically, the BMU, and the neighboring
map vectors within a radius defined  adjust their
values according to a learning principle defined by,
\begin{eqnarray}
\label{learn_1}
V_i(n+1) & = & V_i(n) +  C(n) \alpha(n) [\xi(n) - V_i(n)]  
%\; \; \; i \in N_c(n) \\
%V_i(n+1) & = & V_i(n) \; \; \; i  \!\not\in  N_c(n)
\end{eqnarray}
where $n$ is the iteration number; $\alpha(n)$ is the {\em learning function}; $C(n)$ is the {\em neighborhood function} for the BMU defined as a circle of decreasing radius.
Both  $C(n)$, and $\alpha(n)$, decrease monotonically with $n$, so that an initial global ordering is established at the first iteration, while the subsequent iterations yield more specific adjustments. 
In our case we use square maps of size $L_{MAP}$, and 
\begin{eqnarray}
\label{learn_2.1}
\alpha(n) & = & L_{MAP}  \left(\frac{n_{train} -n}{n_{train}} \right) \\
\label{learn_2.2}
R & = & R_{MAP} \left(\frac{n_{train} -n}{n_{train}} \right) \\
\label{learn_2.3}
C(n) & = & \exp\left(- \frac{D_i^2}{2R^2}\right)
\end{eqnarray}
where in our case, $L_{MAP}=1$, $n_{train} $ is the (variable) total number of iterations, $R_{MAP}$ defines the monotonically decreasing radius, and $K_R=1.5$.

\vspace{0.5cm}
\noindent {\it iii) Mapping} 

\noindent
At the end of a properly trained SOM, cells  that are topologically close to 
each other will contain data which are similar to each other.
In the final phase the actual data gets distributed on the map and 
clusters emerge. 
Since each map vector now
represent a class of similar objects, the SOM
is an ideal tool to
visualize high-dimensional data, by projecting it onto a low-dimensional map
clustered according to some desired similar feature. 
In what follows we apply these ideas to PDF fitting.

%%%%%%%%%%%%%%%%%
\section{SOMPDF Parametrization} 
\label{sec3}
As an illustrative example we fit a subset of all available DIS data. Our fit uses SOMs to generate and classify possible ``candidate" distribution functions.  A Genetic Algorithm  (GA) is subsequently applied.  

In the initialization stage, a set of database/input PDFs is formed by randomly selecting them from a range of functional forms tabulated as functions of $(x,Q^2)$. 
The input vectors could be in principle chosen completely at random. However, in all practical SOM applications, some initial ordering limiting the type and number of 
possible map vectors is given \cite{Kohonen}.      
In our case, the goal is to obtain a stochastic ensemble that still preserves the smoothness of the curves.  The initial PDFs are constructed  to be linear combinations of functions of a similar form to the ones provided by existing global analyses parametrization sets
\cite{CTEQ6,AMP06,CTEQ_2010,GRV}. For each parameterization set we perform random variations on the {\it parameters}, instead than on the functions' values themselves. 
Baryon number and momentum sum rules are imposed at every step.
From our initial analysis we obtained that it is sufficient to combine three different PDF types in order to achive both the required smoothness and variety/randomness. 
This constitutes a considerable improvement over the various initialization procedures tested in \cite{Carnahan} in that  no smoothing techniques are now required.
The input PDFs, are connected through evolution, and convoluted with the pQCD coefficient functions at NLO. 
A subset of the input PDFs is finally placed on the map.

\begin{figure}[h]
\centerline{\includegraphics[width=8cm]{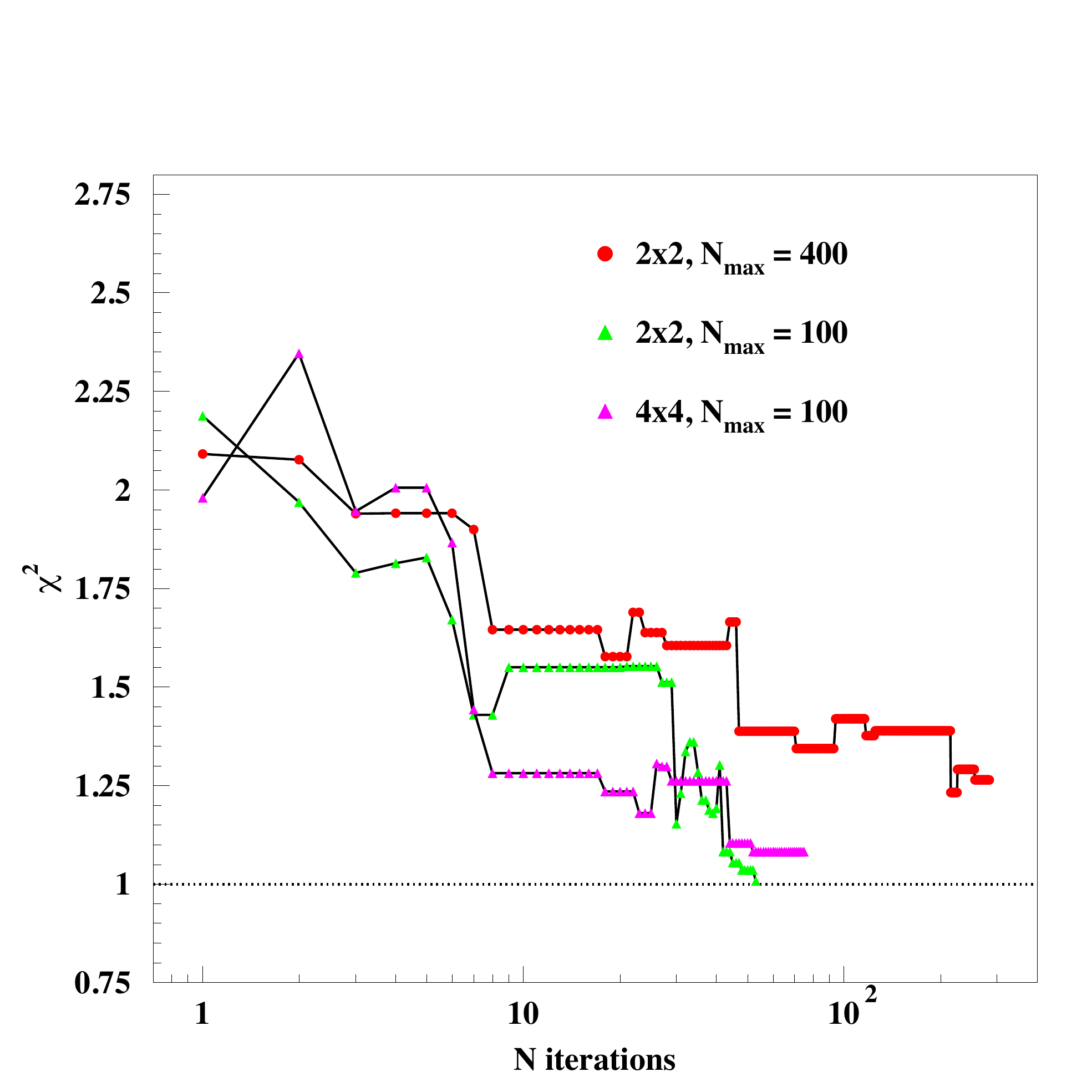}}
%\hspace{2pc}%
%\begin{minipage}[b]{14pc}\caption{\label{label}Figure caption for a narrow figure where the caption is put at the side of the figure.}
%\end{minipage}
\caption{\label{fig1} $\chi^2$ vs. number of iterations in the GA, obtained by fixing the values of the algorithm's parameters such as the maximum number of iterations, the maximum value of $\chi^2$ allowed, and the size of the map. 
}
\end{figure}

For the training stage, a different subset of the input PDFs is used. 
The similarity between our map cell vectors and the code vectors is computed by comparing the PDFs, according to Eqs.(\ref{learn_1},\ref{learn_2.1},\ref{learn_2.2},\ref{learn_2.3}). 
at given $x$ and $Q^2$ values. The new map PDFs are obtained by averaging the neighboring PDFs with the BMU. Once the map is trained, the GA is implemented. The $\chi^2$ per input PDFs is calculated with respect to the experimental data. We then take a subset of these functions with the best  $\chi^2$, and use them as seeds to form a new set of input PDFs. We train the map with the new set of input PDFs and repeat the process. The $\chi^2$ was found to decrease towards $\chi^2=1$ with every GA iteration.  Our stopping criterion is established when the $\chi^2$ stops varying (its curve flattens). In Figure \ref{fig1} we show results on the dependence of the $\chi^2$ on the number of iterations in the GA, obtained using small size maps for illustrative purposes. The role of the different values of the algorithm's parameters can be seen. 
%%% FIGURE 1
%%%
%%% FIGURE 2
\begin{figure}
\centering
\centerline{\includegraphics[width=0.75\textwidth]{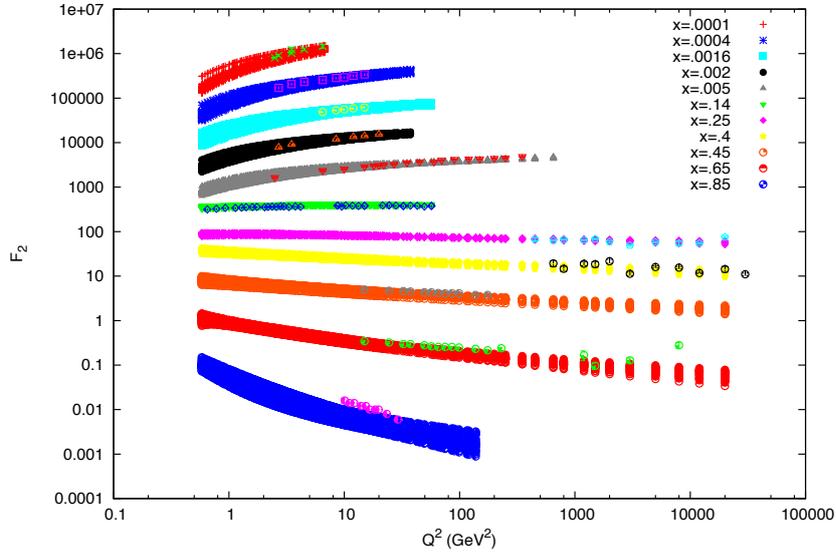}}
\caption{A set of map vectors from the PDF envelope characterizing the SOMPDF initialization stage. Data from SLAC, BCDMS, H1 and ZEUS are shown. }
	\label{fig2}
\end{figure}
%%%
%%% FIGURE 3
\begin{figure}
\centering
\centerline{\includegraphics[width=0.75\textwidth]{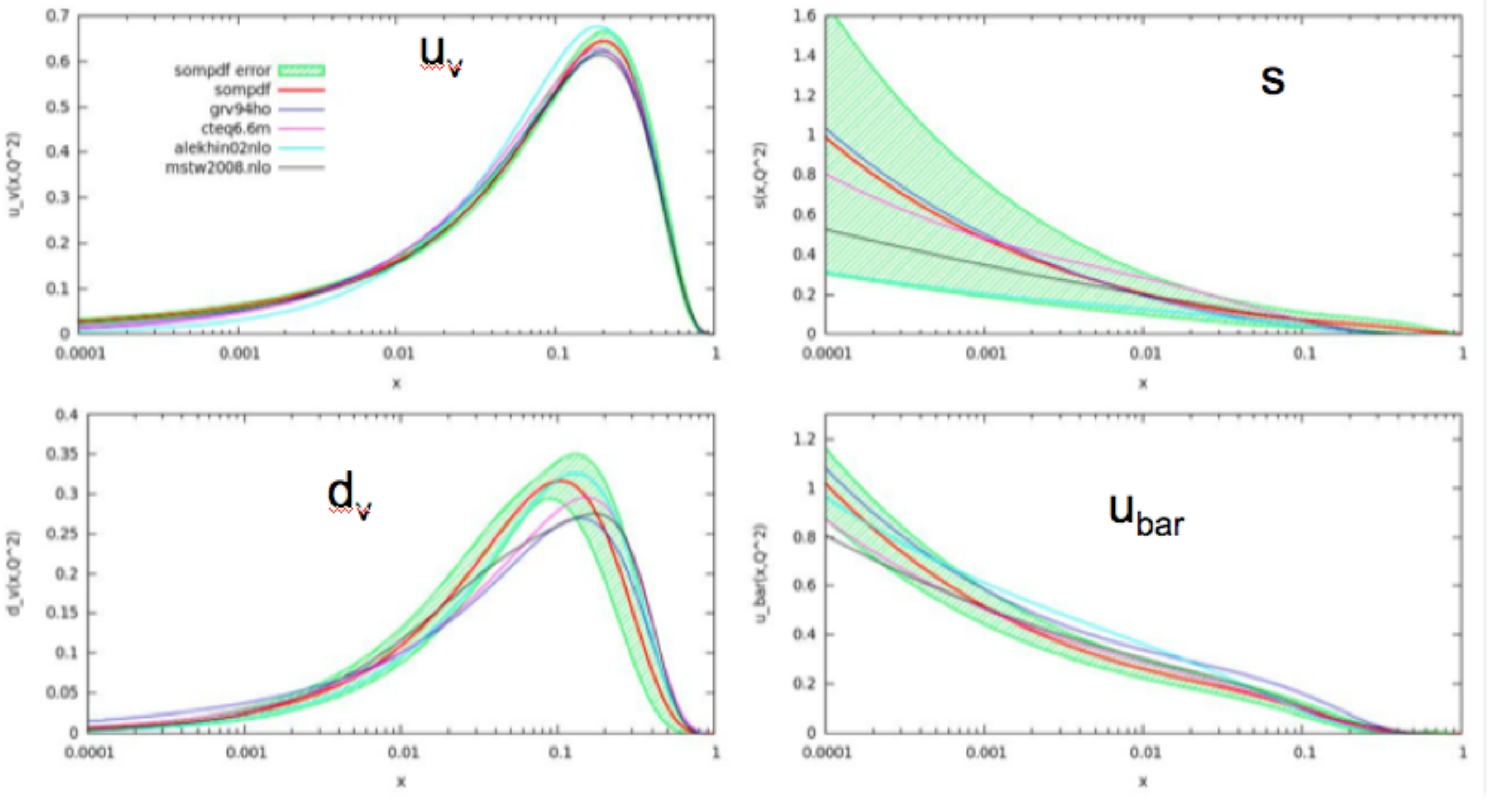}}
\caption{Test results using a $5 \times 5$ map for a set of 43 runs. The panels represent (clockwise from upper left)the $u_v$,
$s$, $d_v$, $\bar{u}$ distributions, respectively, at $Q^2 = 7.5$ GeV$^2$. The shaded areas are our results including the error analys
outlined in the text. For comparison we show also results from several other parametrizations.}
	\label{fig3}
\end{figure}
%%%%%%%%%%%%%%

In Figure \ref{fig2}, we illustrate the shape and size of the envelopes constituting the initial set of map PDFs.   A ``test" set of data from BCDMS + is used. 

Finally, in Figure \ref{fig3} we show results of runs using more complete data sets, consistent with the sets used in Refs. \cite{Ball.2010,CTEQ6,AMP06}. 
The data sets chosen were from BCDMS, H1, NMC, SLAC and ZEUS (see {\it e.g.} references in \cite{DelDebbio.2007}). 
%The following cutoffs were adopted $Q^2 < 1.3$ GeV, $x < .0001$ and $W > 8$ GeV. 

%\begin{figure}[h]
%\begin{minipage}{14pc}
%\includegraphics[width=14pc]{name.eps}
%\caption{\label{label}Figure caption for first of two sided figures.}
%\end{minipage}\hspace{2pc}%
%\begin{minipage}{14pc}
%\includegraphics[width=14pc]{name.eps}
%\caption{\label{label}Figure caption for second of two sided figures.}
%\end{minipage} 
%\end{figure}
%%%%
\section{Conclusions and Future Developments}
%Pioneering work using SOMs for the analysis of  high energy physics data was discussed in \cite{Lonn}.
In Ref.\cite{Carnahan} a new approach, SOMPDF, was presented as an alternative to the purely automated fitting procedure of NNPDF. Initial results
were aimed at proving the viability of the method as a fitting procedure, by successfully generating $\chi^2$ values that decreased
with the number of iterations. These studies did not focus on the specific clustering properties of the map.
The clustering features are, however, important, especially when extending the fitting procedure to a wider set of semi-inclusive and exclusive scattering experimental
data which purport to extract the Transverse Momentum Distributions (TMDs) and Generalized Parton Distributions (GPDs), respectively. 
We believe that SOMPDFs are an ideal tool for analyzing these rather elusive observables because of the following reasons:

%Advantages of SOMs
%Still black box but users' intervention. 
%Data driven
{\it i)}  the extrapolation to regions where data are  scarce is feasible, since the SOM method does not rely on outputs for the learning process; 

{\it ii)} visualization advantages are a key feature of SOMs \cite{Kohonen,Lonn};

{\it iii)} one obtains an improved physical insight on multivariable dependent phenomena. 
This is accomplished by first singling out the main features of the data through the clustering property of the SOM,
%%%%
\footnote{SOMs are in fact also defined as a non linear extension of Principal Component Analysis (PCA).},
%%%%
and by subsequently inspecting the content of the map cells that are sensitive to those features (see {\it e.g.} \cite{Lonn}). 

%The main point is that since for NNPDFs the effect of modifying individual
%NN parameters is unknown, the result might not be under control
%in the extrapolation region, or in between the data points if the data are sparse. 
We therefore refined the SOMPDF approach of  Ref.\cite{Carnahan},  and restructured the original code in such a way that 
on one side, smooth fitting curves are obtained and a fully quantitative error analysis can be now implemented, and on the other we have sufficient flexibility to allow for analyses of different observables, including  the matrix elements for deeply virtual exclusive and semi-inclusive processes.
Our first quantitative results for the unpolarized case using Next-to-Leading-Order (NLO) perturbative QCD in the $\overline{MS}$ scheme were presented in Ref.\cite{Perry_dis10,Hol_exc}, and at this workshop.

Based on our results we are now confident that the SOMPDF method stands as a viable tool for PDF analyses, which is complementary to NNPDF. 
It is also an essential tool to be used in more complex multivariable analyses.

More work is in progress that will refine our initial results, and eventually provide PDFs with an improved error analysis.  
As, in fact, recently articulated in \cite{CTEQ_2010}, the evaluation of the PDFs' uncertainty is complicated because it originates from different experimental and theoretical sources. In particular, the treatment of the theoretical errors cannot be uniquely defined, 
since their correlations are not well known. Our approach is currently defining statistical errors on an ensemble of SOMPDF runs. 
An evaluation using the Lagrange multiplier method is in progress.


\section*{References}
\begin{thebibliography}{9}
\bibitem{Pet}  C.~Peterson, T.~Rognvaldsson, L.~Lonnblad,
  %``JETNET 3.0: A Versatile artificial neural network package,''
  Comput.\ Phys.\ Commun.\  {\bf 81}, 185-220 (1994); C.~Peterson,
  in Proceedings of Workshop on ``{\it New Computing Techniques in Physics Research II} '',  
  Edited by D. Perret-Gallix. River Edge, N.J., World Scientific, 1992. 
  
\bibitem{Ball.2010}
  R.~D.~Ball, L.~Del Debbio, S.~Forte, A.~Guffanti, J.~I.~Latorre, J.~Rojo and M.~Ubiali,
  %``A first unbiased global NLO determination of parton distributions and their
  %uncertainties,''
  Nucl.\ Phys.\  B {\bf 838} (2010) 136;   
  JHEP {\bf 1005}, 075 (2010);
  Nucl.\ Phys.\  B {\bf 823}, 195 (2009);
  Nucl.\ Phys.\ B{\bf 809}, 1 (2009), Erratum-ibid. B{\bf 816}, 293 (2009).
  
\bibitem{DelDebbio.2007}
  L.~Del Debbio, S.~Forte, J.~I.~Latorre, A.~Piccione and J.~Rojo  [NNPDF
                  Collaboration],
  %``Neural network determination of parton distributions: the nonsinglet
  %case,''
  JHEP {\bf 0703}, 039 (2007); {\it ibid} JHEP {\bf 0503}, 080, (2005).

\bibitem{Forte.2002}
  S.~Forte, L.~Garrido, J.~I.~Latorre and A.~Piccione,
  %``Neural network parametrization of deep-inelastic structure functions,''
  JHEP {\bf 0205}, 062 (2002).

\bibitem{Kohonen} T. Kohonen, Self-organizing Maps (Springer, New York, 2001), 3rd. ed.
  
\bibitem{Carnahan} H. Honkanen, S. Liuti, J. Carnahan, Y. Loitiere, P. R. Reynolds, Phys. Rev. D 79, 034022 (2009)

\bibitem{Perry_dis10} D.~Z.~Perry, K.~Holcomb and S.~Liuti,
  arXiv:1008.2137 [hep-ph].
  
\bibitem{Hol_exc}  K.~Holcomb, S.~Liuti and D.~Z.~Perry, 

\bibitem{CTEQ6} J. Pumplin, D. R. Stump, J. Huston, H. L. Lai, P. Nadolsky, W. K. Tung (CTEQ6), Phys.\ Rev.\  D {\bf 65}, 014013 (2001)
 
\bibitem{AMP06} S. Alekhin, K. Melnikov, F. Petriello (AMP06), Phys. Rev. D 74, 054033 (2006) 

\bibitem{CTEQ_2010} H.~L.~Lai, J.~Huston, Z.~Li, P.~Nadolsky, J.~Pumplin, D.~Stump and C.~P.~Yuan,
  arXiv:1004.4624 [hep-ph].  

\bibitem{GRV} M.~Gluck, E.~Reya and A.~Vogt,
  Eur.\ Phys.\ J.\ C\ {\bf 5}, 461  (1998)

\bibitem{Lonn} L.~Lonnblad, C.~Peterson, H.~Pi, T.~Rognvaldsson,
  %``Selforganizing networks for extracting jet features,''
  Comput.\ Phys.\ Commun.\  {\bf 67}, 193-209 (1991).

\end{thebibliography}
\end{document}